\documentclass[preprint,12pt]{elsarticle}

\usepackage[utf8]{inputenc}
\usepackage[english]{babel}
\usepackage{graphicx}
\usepackage{xcolor}
\usepackage{amsmath, amsfonts, dsfont}
\usepackage{natbib}
\usepackage{diagbox}
\usepackage{algorithm2e}
\usepackage{algorithm2e}
\usepackage{url}

\usepackage{placeins}

\begin{document}

\begin{frontmatter}

\title{Examining the robustness of a model selection procedure in the binary latent block model \\ through a language placement test data set}
\author{Vincent Brault, Fr\'ed\'erique Letu\'e, Marie-Jos\'e Martinez}

This preprint has not undergone peer review (when applicable) or any post-submission improvements or corrections. The Version of Record of this article is published in Statistical Methods and Applications, and is available online at \url{https://doi.org/10.1007/s10260-026-00844-1}.

\address{Univ. Grenoble Alpes, CNRS, Grenoble INP\footnote{Institute of Engineering Univ. Grenoble Alpes}, LJK, 38000 Grenoble, France}

\begin{abstract}
When entering French university, the students' foreign language level is assessed through a placement test. In this work, we model the placement test results using binary latent block models which allow to simultaneously form homogeneous groups of students and of items. However, a major difficulty in latent block models is to select correctly the number of groups of rows and the number of groups of columns. The first purpose of this paper is to tune the number of initializations needed to limit the initial values problem in the estimation algorithm in order to propose a model selection procedure in the placement test context. Computational studies based on simulated data sets and on two placement test data sets are investigated. The second purpose is to investigate the robustness of the proposed model selection procedure in terms of stability of the students groups when the number of students varies.
\end{abstract}


\end{frontmatter}

\section{Introduction}

When entering French university, students may have different foreign language levels. They need to be evaluated before being directed in an adequate foreign language class. To assess their level, universities use different placement tests. The SELF test developed in Grenoble is one of the most used (\cite{cervini, brault2021}). At the end of this test, each student get an aggregated score which corresponds to the course level where he/she has to register in. He/she also gets a mark for each of the three evaluated skills (oral comprehension, written comprehension and written expression). Moreover, the test creators need to check the relevance of the questions, in particular whether they are discriminating or not. It may be useful to form groups of items more or less difficult per skill.

\medskip
Results of such a placement test can be displayed as a matrix where each row corresponds to one student and each column to one item. Element $(i,j)$ of the matrix equals 1 if student $i$ answers correctly question $j$, and 0 otherwise. In their native paper, Govaert and Nadif \cite{govaert2003} propose latent block models (\cite{govaert2003}) which simultaneously achieve a clustering of the rows and the columns and thus turn out to be particularly useful to form homogeneous groups of students and of items in the placement test context. In their paper, they present simulation studies but no theoretical results. Only in 2015, Keribin et al. \cite{keribin2015} give sufficient conditions for the identifiability of the latent block model for binary data.   

\medskip
In latent block models, maximum likelihood estimators are not computationally available. That is the reason why algorithms extending the EM-algorithm (\cite{dempster}) to estimate parameters, namely Variational or Stochastic Expectation Maximization, have been proposed in \cite{govaert2003,govaert2008,keribin2015}. Results on asymptotic convergence of maximum likelihood and variational estimators are only been obtained in 2020 by Brault et al. \cite{brault2020consistency}. In practice, although these algorithms can give satisfactory estimates, they appear to be quite sensitive to starting values and have a marked tendency to provide empty clusters. To overcome these limitations, Keribin et al. \cite{keribin2015} proposed several algorithms through Bayesian inference using Gibbs sampling. 

\medskip
A major difficulty in latent block models is to select correctly the number of groups of rows and the number of groups of columns. For this purpose, penalized likelihood criteria such as Akaike Information Criterion (AIC, \cite{akaike1973,akaike1974}) or Bayesian Information Criterion (BIC, \cite{schwarz1978}) are not directly usable since computing the maximized likelihood is not possible. Another widely used criterion is the Integrated Completed Likelihood criterion (ICL) defined by Biernacki et al. \cite{biernacki2000}. This criterion has been extended to the latent block model for binary data in \cite{keribin2012} and for categorical data in \cite{keribin2015}. Note that, to our knowledge, no theoretical result nor non-asymptotic concentration inequalities on these model selection criteria are available. 

\medskip
In this work, we are faced with similar theoretical difficulties. Therefore, we adopt in this paper a computational approach. Our first purpose is to tune the number of initializations needed to limit the initial values problem in the estimation algorithm in order to propose a model selection procedure for latent block models on binary data in the context of placement tests. Two computational studies based on simulated data sets and on the two placement test data sets described in Section \ref{sec:datasets} are investigated. Our second purpose is to investigate the robustness of the proposed model selection procedure. The robustness is here assessed in terms of the stability of the students groups when the number of students varies.

\medskip
The paper is organized as follows. The two placement test data sets used in this paper are described in the next section. In Section \ref{sec:framework}, we recall the statistical model, the notations and the estimation algorithm and we present the model selection procedure based on ICL criterion considered in this work. Section \ref{sec:initstrat} deals with the effect of initialization strategies on the model selection procedure. The robustness of the proposed procedure is investigated in Section \ref{sec:robustness}. Finally, a conclusion presenting a perspective for future work ends this paper.

\section{Data sets}\label{sec:datasets}

The data sets considered in this paper have been obtained from the SELF placement test developed at Université Grenoble Alpes in 6 languages (English, French as a Foreign Language, Italian, Japanese, Mandarin Chinese and Spanish) and used at a number of partner universities in France (see \cite{cervini}). The SELF test is a semi-adaptive multi-stage test. The first stage (the initial testlet) is common to all test takers, but the items in the second stage depend on test takers’ results in the first. Results in the second stage are used to refine the estimation of learners’ level and arrive at placement results expressed in Common European Framework of Reference (CEFR) levels that are as reliable as possible. In this paper, the results of a Japanese and an English SELF placement tests are considered. The Japanese (resp. English) SELF test has been taken by 137 (resp. 228) students entering Université Grenoble Alpes in 2019. We focus here only on the first stage composed with 33 (resp. 36) items common to all test takers. These items are labelled with three skills: oral comprehension, written comprehension and written expression. 

\begin{figure}
\begin{center}
    \begin{tabular}{cc}
        \begin{minipage}{0.45\linewidth}
        \includegraphics[scale=0.4]{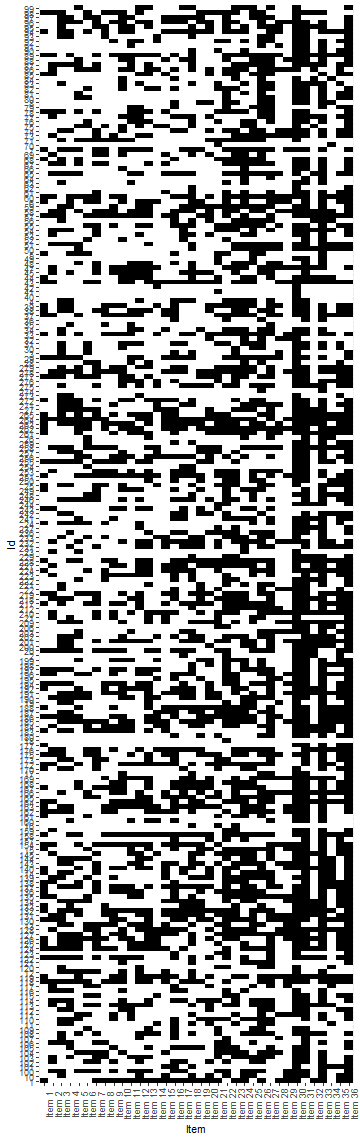}
        \end{minipage} &
        \begin{minipage}{0.45\linewidth}
        \includegraphics[scale=0.4]{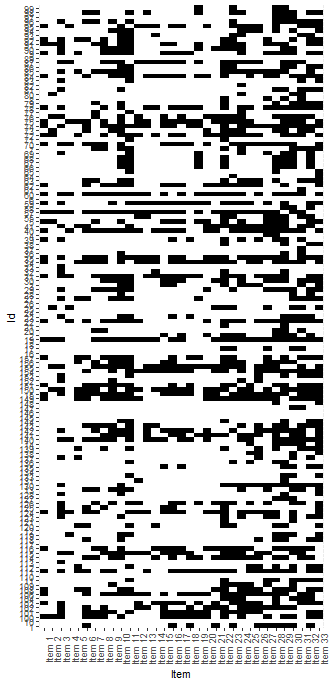}
        \end{minipage}  \\
    \end{tabular}
    \caption{Results of the English SELF placement test (on left) for 228 students (lines) and 36 items (columns) and the Japanese SELF placement test (on right) for 137 students (lines) and 33 items (columns). A white cell corresponds to a correct answer.\label{donnees_brutes}}
\end{center}
\end{figure}

\medskip
Figure \ref{donnees_brutes} displays the English SELF placement test results for the 228 students (lines) and the 36 items (columns). Each element of the matrix is colored in white when the student answers correctly the question, and in black otherwise.

\section{Statistical framework}\label{sec:framework}

In this section, the statistical model, the notations and the estimation algorithm are first described. Then, the model selection procedure based on ICL criterion is presented.

\subsection{Statistical model and notations}\label{Statmod}

Let $n$ be the number of students, $g$ the number of students groups, $q$ the number of items and $m$ the number of items groups. We define $Z_i, i=1 \cdots n$, as the independent random variables modelling the students group which student $i$ belongs to. The $Z_i, i=1 \cdots n$ variables follow the multinomial $\mathcal{M}(1; \pi_1, \dots, \pi_g)$ distribution. We also define $W_j, j=1 \cdots q$, as the independent random variables modelling the items group which item $j$ belongs to. The $W_j, j=1 \cdots q$ variables follow the multinomial $\mathcal{M}(1; \rho_1, \dots, \rho_m)$ distribution.     

\medskip
Responses are assumed to be independent. Given that student $i$ belongs to group $k$ and item $j$ belongs to group $l$, the response of student $i$ to item $j$ denoted by $Y_{ij}$ follows the Bernoulli distribution with parameter $\alpha_{kl}$:

$$P(Y_{ij}=1| Z_i=k, W_j=l) = \alpha_{kl}.$$

\subsection{Estimation algorithm}\label{sec:estimation}
A classical method to estimate parameters in latent variables models is the EM-algorithm. \cite{govaert2003,govaert2008} show that the EM-algorithm cannot directly be used in practice. Indeed, it would require to calculate a sum over all possible couples $(z_i,w_j), 1 \leq i \leq n, 1 \leq j \leq q$, which would have a too high computational cost. To overcome this problem, \cite{govaert2008} propose a variational approximation of the EM-algorithm (VEM) based on a decomposition of the log-likelihood as the sum of the free energy and the Kullback-Leibler divergence (\cite{kullback}). Since the Kullback-Leibler divergence is expected to be small around the log-likelihood maximum, they propose to only maximise the free energy. In \cite{brault2020consistency}, the authors show that the free energy maximum estimator is consistent.

\medskip
 As pointed by Govaert and Nadif \cite{govaert2008} and Keribin et al. \cite{keribin2015}, the estimated parameters obtained by the VEM-algorithm highly depend on its initial values. Moreover, Keribin et al. \cite{keribin2015} show that this algorithm tends to provide empty groups. To overcome these problems, they propose a V-Bayes algorithm to avoid empty groups, combined with a Gibbs sampler to limit the initial values problem. In this bayesian approach, proper and independent informative prior distributions are considered for the parameters. The mixing proportions $\pi=(\pi_1, \dots, \pi_g)$ and $\rho=(\rho_1, \dots, \rho_m)$ are assumed to be Dirichlet-distributed with parameters $(a,\dots,a)$. The parameters $\alpha_{11}, \dots, \alpha_{gm}$ are assumed to be Beta-distributed with parameter $(b,b)$. In Section \ref{sec:initstrat}, hyperparameters $a$ and $b$ in the estimation algorithm are chosen to be equal to 4 and 1 respectively, as advised by Keribin et al. \cite{keribin2015}. In this paper, we use the combined V-Bayes Gibbs algorithm to get estimators $\widehat{\pi}, \widehat{\rho}, \widehat{\alpha}$ of the parameters for a given number of students groups $g$ and a given number of items groups $m$. Furthermore, in order to limit the initial values problem, this algorithm is run $T$ times and we finally keep the estimators $\widehat{\pi}, \widehat{\rho}, \widehat{\alpha}$ which maximize the free energy over the $T$ runs.

\subsection{Model selection procedure and the ICL criterion}
In this subsection, we investigate the ICL model selection criterion in order to propose a model selection procedure. 

\medskip
The ICL criterion has been defined in \cite{biernacki2000} as the logarithm of the integrated completed likelihood in a mixture model context. It has been extended to the latent block model for binary data in \cite{keribin2012} and for categorical data in \cite{keribin2015}.

\medskip
Using the conditional independence of the $z$'s and the $w$'s conditionally to $\pi$, $\rho$ and $\alpha$, the conjugate properties of the prior Dirichlet distributions and the conditional independence of the $y_{ij}$ given the latent variables $z$ and $w$, the ICL criterion can be written as 

\begin{eqnarray*}
ICL_{(z,w)}(g,m) & = & \log \Gamma(ga) + \log \Gamma(ma) - (m + g) \log\Gamma(a) \\
                & & + mg(\log \Gamma(2b) - 2 \log \Gamma(b)) - \log \Gamma(n + ga) - \log \Gamma(q + ma)\\
                & & +\sum_k \log \Gamma(z_{.k} + a) + \sum_l \log\Gamma(w_{.l} + a)\\
                & & + \sum_{k,l} \Bigg[  \log \Gamma(N^1_{kl}+b) + \log \Gamma(N^0_{kl}+b) - \log \Gamma(z_{.k} w_{.l} + 2b) \Bigg]
\end{eqnarray*}
where 
\begin{eqnarray*}z_{.k}= \sum_i \mathds{1}_{\{z_i=k\}},& & w_{.l}= \sum_j \mathds{1}_{\{w_j=l\}},\\ N^1_{kl}=\sum_{i,j} y_{ij} \mathds{1}_{\{z_i=k, w_j=l\}}, & &  N^0_{kl}=\sum_{i,j} (1-y_{ij}) \mathds{1}_{\{z_i=k, w_j=l\}}.
\end{eqnarray*} 
Details can be found in \cite{keribin2015}.

\medskip
Given $(g,m)$, the maximization of $ICL_{(z,w)}(g,m)$ over all partitions $(z,w)$ would require a too high computational cost. In their paper, Biernacki et al. {biernacki2000} proposed to use $ICL_{(\widehat{z},\widehat{w})}(g,m)$ where $(\widehat{z},\widehat{w})$ is the maximizing partition obtained with a maximum a posteriori (MAP) rule after the last V-Bayes step.

\medskip
In order to select the numbers of students groups $g$ and of items groups $m$, we calculate every criterion $ICL_{(\widehat{z},\widehat{w})}(g,m)$ where $(g,m)$ runs on a given grid, and we finally define $(\widehat{g},\widehat{m})$ as the pair that maximizes these criteria.

\section{Effect of initialization strategies on the model selection procedure}\label{sec:initstrat}

The objective of this section is to tune the parameter $T$, namely the number of initializations needed to limit the initial values problem in the estimation algorithm. For that purpose, we investigate two computational studies. The first one is based on simulated data sets, and the second one is based on the two real data sets described in Section \ref{sec:datasets}.

\subsection{Simulation study}
\subsubsection{Simulation plan and indicators}\label{simu}

In this simulation study, we simulate $L= 100$ data sets from the model described in Section \ref{Statmod}. Mimicking the Japanese SELF placement test data set, we set $n= 137$ and $q=33$. We also set $g=3$, $m=4$, $\pi_1=\pi_2=\pi_3=\frac{1}{3}$ and $\rho_1=\rho_2=\rho_3=\rho_4=\frac{1}{4}$. Finally, we define parameter $\alpha_{kl}$ equal to $\varepsilon$ if $k \geq l$ and $1-\varepsilon$ if $k < l$ with $\varepsilon \in \{0.05, 0.15, 0.2, 0.25, 0.3\}$. Note that parameter $\varepsilon$ can be considered as an indicator of the estimation difficulty.

\medskip
In order to tune parameter $T$, we run the following algorithm:

\medskip
\begin{algorithm}
\For{$\varepsilon \in \{0.05, 0.15, 0.2, 0.25, 0.3\}$}{
    \For{$L$ from 1 to 100}{
        simulate a data set 
        
        $T=0$ 
        
        $(\hat{g}, \hat{m})=(1,1)$
        
        \While{$(\hat{g}, \hat{m}) \neq (3,4)$}{
            $T = T+1$
            
            \For{each pair $(g,m)$, with $g$ and $m$ varying from 1 to 7}{
                \begin{enumerate}
                \item calculate the estimators $(\widehat{\pi}, \widehat{\rho}, \widehat{\alpha})$;
                \item calculate the maximizing a posteriori partition $(\widehat{z},\widehat{w})$;
                \item calculate the associated ICL value;
                \end{enumerate}
                }
            select the pair $(\hat{g}, \hat{m})$ that maximizes the ICL criterion.
        }
    }
}
\end{algorithm}

\medskip
Table \ref{tab:T_varia_sans_mediane} displays the distribution of parameter $T$ for each value of $\varepsilon$.

\begin{table}[!h]
    \centering
    \caption{Distribution of $T$ (columns) in function of the value $\varepsilon$ (rows). 
    }
    \label{tab:T_varia_sans_mediane}
    \begin{tabular}{|c||*{4}{c|}}
    \cline{2-5}
    \multicolumn{1}{c|}{}&1&2&14&$\geq50$\\\hline
\hline$\varepsilon=0.05$&96&4&\textcolor{gray!50!white}{0}&\textcolor{gray!50!white}{0}\\
\hline$\varepsilon=0.15$&99&1&\textcolor{gray!50!white}{0}&\textcolor{gray!50!white}{0}\\
\hline$\varepsilon=0.2$&98&\textcolor{gray!50!white}{0}&\textcolor{gray!50!white}{0}&2\\
\hline$\varepsilon=0.25$&79&\textcolor{gray!50!white}{0}&1&20\\
\hline$\varepsilon=0.3$&24&\textcolor{gray!50!white}{0}&\textcolor{gray!50!white}{0}&76\\\hline
    \end{tabular}
\end{table}

\subsubsection{Results}

Results displayed in Table \ref{tab:T_varia_sans_mediane} show that for small values of $\varepsilon$ ($\varepsilon \leq 0.25$), $T=1$ is enough to select the simulated model ($g=3$ and $m=4$) most of the time (96\% for $\varepsilon= 0.05$, 99\% for $\varepsilon= 0.15$, 98\% for $\varepsilon = 0.2$ and 79\% for $\varepsilon = 0.25$). That means that for simple cases running the estimation algorithm once per pair $(g,m)$ is usually sufficient. 

\medskip
For $\varepsilon = 0.3$, we can see from Table \ref{tab:T_varia_sans_mediane} that the distribution of parameter $T$ is shifted to higher values. That means that for more difficult cases, we need to run the estimation algorithm a high number of times per pair $(g,m)$ to retrieve the simulated model.

\subsection{Study based on real data sets}

In this subsection, we evaluate the performance of our procedure from the two data sets described in Section \ref{sec:datasets}. For that purpose, we proceed in two steps. 

\medskip
First, since there is no "true" model in a real dataset context, we run the following algorithm a great number of times $K=170\,000$ to get a reference model :

\medskip
\begin{algorithm}
\For{$k=1$ to $K$}{
    \For{each pair $(g,m)$, with $g$ and $m$ varying from 1 to 7}{
        \begin{enumerate}
            \item calculate the estimators $(\widehat{\pi}, \widehat{\rho}, \widehat{\alpha})$;
            \item calculate the maximizing a posteriori partition $(\widehat{z},\widehat{w})$;
            \item calculate the associated ICL value;
        \end{enumerate}
    }
    $(\hat{g}, \hat{m})_k$ is the pair $(\hat{g}, \hat{m})$ that maximizes the ICL criterion.
}
$(g, m)_{ref}$ is the pair $(\hat{g}, \hat{m})_k$ that maximizes the ICL criterion over the $K$ pairs.
\end{algorithm}

\medskip
Secondly, we consider the number of times this algorithm finds the reference model over the $K$ runs and we study the inter-arrival times of the reference model along the $K$ runs. If these inter-arrival times are quite small, we could consider that the dataset is an "easy" case. The larger inter-arrival times are, the more difficult the case is considered. 

\subsubsection{The Japanese SELF placement test data set}

For the Japanese SELF dataset, the obtained reference model is $(3,4)$. Over the $K=170\,000$ runs, the reference model has been obtained $170\,000$ times. That means that the inter-arrival times are all equal to 1. This dataset can be considered as an "easy" case, for which only one initialization is needed.

\begin{figure}[!ht]
    \centering
    \begin{tabular}{cc}
    \begin{minipage}{0.45\linewidth}
         \includegraphics[scale=0.4]{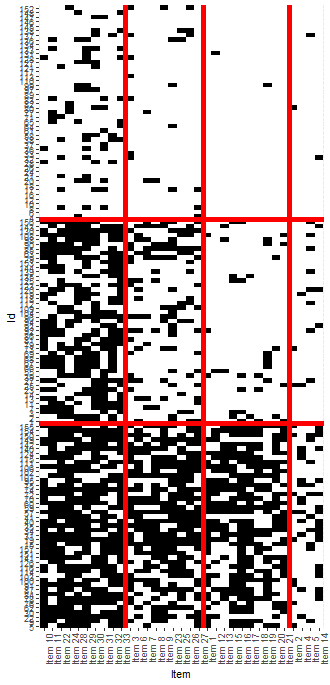}
    \end{minipage}& 
    \begin{minipage}{0.45\linewidth}
         \includegraphics[scale=0.4]{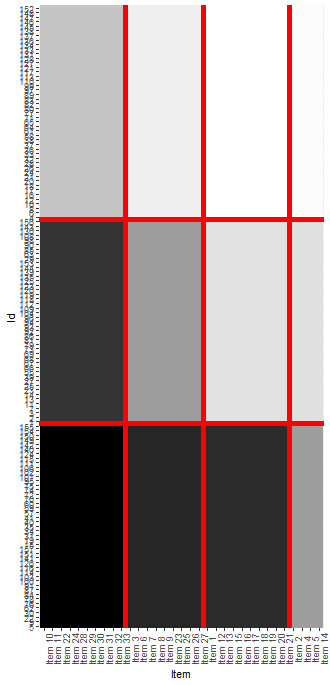}
    \end{minipage} \\
    \end{tabular}
    \caption{Data representation after the rows and columns have been ordered by classes (left) and estimated model representation (right) for the Japanese SELF test data set. The closer the $\hat{\alpha}_{k\ell}$ value is to 1, the whiter the block.\label{fig:jap:organisation_matrix}}
\end{figure}

\medskip
Table~\ref{tab:jap:parameters} gives one realization of the estimation of each parameter and  Figure~\ref{fig:jap:organisation_matrix} displays a summarized representation of the parameters.

\begin{table}[!ht]
    \caption{A realization of the estimation of $\boldsymbol{\rho}$ (top), $\boldsymbol{\pi}$ (left) and $\boldsymbol{\alpha}$ (bottom right) for the Japanese data set.\label{tab:jap:parameters}}
    \[\begin{array}{|*{6}{c|}}
\cline{3-6}
\multicolumn{2}{c|}{}&0.291&0.264&0.299&0.147\\
\cline{3-6}
\multicolumn{6}{c}{}\\
\cline{1-1}\cline{3-6}
0.329&&0.845&0.952&0.995&0.987\\
\cline{1-1}\cline{3-6}
0.342&&0.414&0.736&0.919&0.916\\
\cline{1-1}\cline{3-6}
0.329&&0.259&0.37&0.388&0.738\\
\cline{1-1}\cline{3-6}
\end{array}\]

\end{table}

\subsubsection{The English SELF placement test data set}

For the English SELF test data set, the obtained reference model is $(4,5)$. This reference model has been obtained only $16$ times over the $K=170\,000$ runs. Thus, we observe 16 inter-arrival times. Table~\ref{tab:eng:distri:T_opt} displays the distribution of these 16 values. They vary between $700$ and $36\,345$, with a median value equal to $6595.5$. This means that we should run at least 6600 initializations to obtain the reference model with probability $1/2$. 

\begin{table}[!ht]
    \centering
    \caption{Distribution of the inter-arrival times of the reference model $(4,5)$ for the English SELF dataset. \label{tab:eng:distri:T_opt}}
    \begin{tabular}{|c|c|}
    \hline
    Max.&36345\\\hline
3rd Qu.&13398.5\\\hline
Mean&10534.125\\\hline
Median&6595.5\\\hline
1st Qu.&4533.75\\\hline
Min.&700\\\hline

    \end{tabular}
\end{table}

\begin{figure}[!ht]
    \centering
    \begin{tabular}{cc}
    \begin{minipage}{0.45\linewidth}
         \includegraphics[scale=0.4]{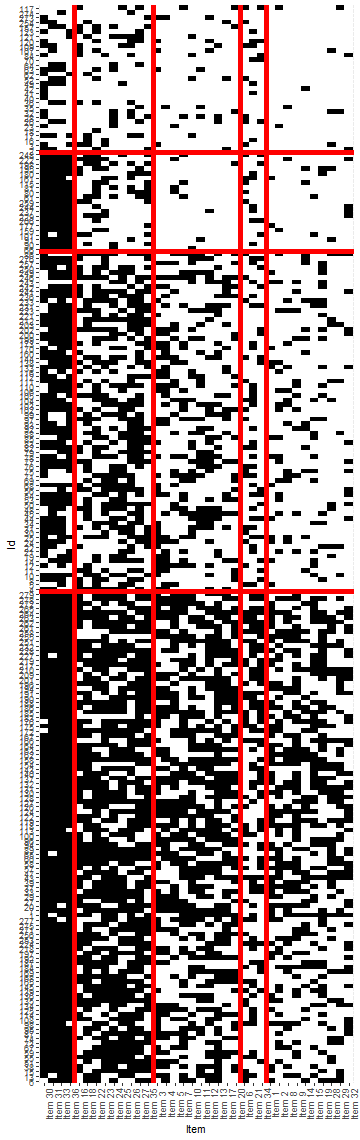}
    \end{minipage}& 
    \begin{minipage}{0.45\linewidth}
         \includegraphics[scale=0.4]{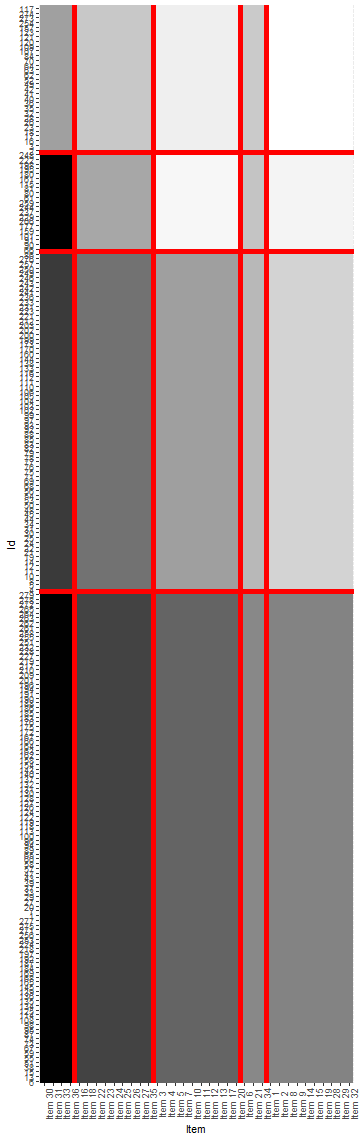}
    \end{minipage} \\
    \end{tabular}
    \caption{Data representation after the rows and columns have been ordered by classes (left) and estimated model representation (right) for the English SELF test data set. The closer the $\hat{\alpha}_{k\ell}$ value is to 1, the whiter the block.\label{fig:eng:organisation_matrix}}
\end{figure}

Table~\ref{tab:eng:parameters} gives one realization of the estimation of each parameter and  Figure~\ref{fig:eng:organisation_matrix} displays a summarized representation of the parameters.

\begin{table}[!ht]
    \caption{A realization of the estimation of $\boldsymbol{\rho}$ (on top), $\boldsymbol{\pi}$ (on left) and $\boldsymbol{\alpha}$ (on bottom right) for the English SELF test data set.\label{tab:eng:parameters}}
    \[\begin{array}{|*{7}{c|}}
\cline{3-7}
\multicolumn{2}{c|}{}&0.137&0.235&0.256&0.117&0.254\\
\cline{3-7}
\multicolumn{7}{c}{}\\
\cline{1-1}\cline{3-7}
0.145&&0.641&0.779&0.905&0.784&0.958\\
\cline{1-1}\cline{3-7}
0.106&&0.0303&0.664&0.931&0.766&0.922\\
\cline{1-1}\cline{3-7}
0.317&&0.257&0.476&0.637&0.724&0.816\\
\cline{1-1}\cline{3-7}
0.432&&0.0334&0.295&0.424&0.556&0.538\\
\cline{1-1}\cline{3-7}
\end{array}\]

\end{table}

\subsection{Discussion}

The simulation study carried out previously shows that the results can be unstable depending on the difficulty of the considered case. That is why, in practice, we encourage to first run the model selection procedure with only one initialization in the estimation algorithm ($T=1$) and then examine the estimated values $\widehat{\alpha}_{kl}$ of the parameters $\alpha_{kl}$ in the selected model. If the $\widehat{\alpha}_{kl}$ matrix lines and columns have different profiles, we can consider that the case is quite simple and there is no need to increase the number of initializations in the estimation algorithm. This is for instance the case in the Japanese SELF placement test data set (see Figure \ref{fig:jap:organisation_matrix}). On the contrary, when the profiles are similar, it would be relevant to increase the number of initializations in order to stabilize the procedure. As an illustration, one can see in Figure \ref{fig:eng:organisation_matrix} that columns 3 and 5 show quite similar profiles which may explain the selection model difficulties. To determine the relevant number of initializations, a possibility could be to run a simulation study mimicking the real data set from the estimated parameters in order to examine the results stability with respect to $T$.

\section{Robustness}\label{sec:robustness}

This section is devoted to a robustness study of the proposed model selection procedure in the following senses:
\begin{itemize}
    \item the number of students groups with respect to the sample size,
    \item the belonging of two given students to a same group with respect to the sample size.
\end{itemize}

\subsection{Sampling plan and indicators}

To explore the robustness of the proposed model selection procedure, we simulate $L=100$ data sets from the simulation plan described in Section \ref{simu} with $\varepsilon = 0.15$, $0.2$ and $0.25$. For $L=1$ to $100$, we run the algorithm once and we check that $(\hat{g},\hat{m}) =(3,4)$. If this is not the case, we simulate another data set. While running the algorithm, we get the estimated students groups proportions $\widehat{\pi}_1$, $\widehat{\pi}_2$ and $\widehat{\pi}_3$. 

\medskip
For a given students sample size ($n=20, 40, 60,\dots, 120$), we draw $10$ students samples from the $137$ students respecting the $\widehat{\pi}_1$, $\widehat{\pi}_2$ and $\widehat{\pi}_3$ proportions and we apply our procedure to these $10$ samples. We display in Tables~\ref{tab:Evol15}, \ref{tab:Evol20} and \ref{tab:Evol25}, for each value of $\varepsilon$, the distribution of the $(\widehat{g},\widehat{m})$ pairs selected by the proposed model selection procedure with respect to $n$ over the $100 \times 10$ samples.

\medskip
In a second step, we compare the $n$-students partition with the initial $137$-students partition. For that purpose, following \cite{robert} extending \cite{lomet2012},
\begin{itemize}
\item when the selected number of students groups is equal to $g=3$, we draw the contingency table of the students belonging groups in the reference model and in the selected model. Students on the diagonal are defined as well classified whereas the students out of the diagonal are defined as misclassified. As illustrated in Table \ref{table-switching}, we consider all possible labels switching and we keep the one that gives the smallest misclassified students number.
\item when the selected number of students groups is greater than $g=3$, we consider all possible groups unions of the selected model in order to get only $3$ groups and we keep the group union that gives the smaller misclassified students number,
\item when the selected number of students groups is smaller than $g=3$, we consider all possible groups unions of the reference model and we proceed as previously.
\end{itemize}

\begin{table}[h!]
    \centering
    \caption{Example of label switching (\cite{robert}). In the right table, the number of misclassified students is equal to 13. By switching groups $\widehat{G2}$ and $\widehat{G3}$, we obtained a lower number of misclassified students.}
    \null
    \begin{tabular}{c|ccc|c}
    \diagbox[height=2em]{Ref}{} & $\widehat{G1}$ & $\widehat{G2}$ & $\widehat{G3}$& Total \\
    \hline
      G1  & 6 & 1 &  1 & 8\\
      G2  & 0 & 1 &  6 & 7 \\ 
      G3  & 0 & 5 &  0 & 5  \\ 
      \hline
     Total  & 6 & 7 & 7 & 20  \\  
    \end{tabular}
    \hfill
    \begin{tabular}{c|ccc|c}
    \diagbox[height=2em]{Ref}{} & $\widehat{G1}$ & $\widehat{G3}$ & $\widehat{G2}$& Total \\
    \hline
      G1  & 6 & 1 &  1 & 8\\
      G2  & 0 & 6 &  1 & 7 \\ 
      G3  & 0 & 0 &  5 & 5  \\ 
      \hline
     Total  & 6 & 7 & 7 & 20  \\  
    \end{tabular}
     \label{table-switching}
\end{table}

Tables~\ref{tab:Evol15}, \ref{tab:Evol20} and \ref{tab:Evol25} display the distribution of $(\widehat{g},\widehat{m})$ pairs selected by the proposed procedure with respect to $\varepsilon$ and $n$. We can observe that, for $\varepsilon = 0.15$, the distribution of $(\widehat{g},\widehat{m})$ is very well concentrated on the reference pair $(3,4)$, even when the number of students $n$ is small. As $\varepsilon$ increases, the distribution is more scattered, mainly for small $n$ values. Nevertheless, for $\varepsilon=0.25$, when $n$ increases, we retrieve a concentrated distribution around the reference pair.

\begin{table}[h!]
    \centering
    \caption{Distribution of the $(\widehat{g},\widehat{m})$ pairs selected by the proposed procedure with respect to $n$ for $\varepsilon=0.15$. The boxed data corresponds to the reference pair.}
    \null
   \label{tab:Evol15}
  \begin{tabular}{c|c} 
  \multicolumn{2}{c}{$\varepsilon=0.15$}\\
  \hline\hline
    $n=20$&
    $n=80$\\
         \begin{minipage}{0.45\linewidth}
         \centering
       \scriptsize
         \begin{tabular}{c|cccccc|c}
         \backslashbox{g}{m}&3&4&5&6&Total\\\hline
2&\textcolor{red!19!blue}{190}&\textcolor{red!1.3!blue}{13}&\textcolor{black!20!white}{0}&\textcolor{black!20!white}{0}&\textcolor{red!20.3!blue}{203}\\\cline{3-3}
3&\textcolor{red!4.5!blue}{45}&\multicolumn{1}{|c|}{\textcolor{red!74!blue}{740}}&\textcolor{red!0.2!blue}{2}&\textcolor{black!20!white}{0}&\textcolor{red!78.7!blue}{787}\\\cline{3-3}
4&\textcolor{red!0.9!blue}{9}&\textcolor{red!0.1!blue}{1}&\textcolor{black!20!white}{0}&\textcolor{black!20!white}{0}&\textcolor{red!1!blue}{10}\\

         \end{tabular}
         \end{minipage}&
         \begin{minipage}{0.45\linewidth}
         \centering
       \scriptsize
         \begin{tabular}{c|cccccc|c}
         \backslashbox{g}{m}&3&4&5&6&Total\\\hline
2&\textcolor{black!20!white}{0}&\textcolor{black!20!white}{0}&\textcolor{black!20!white}{0}&\textcolor{black!20!white}{0}&\textcolor{black!20!white}{0}\\\cline{3-3}
3&\textcolor{black!20!white}{0}&\multicolumn{1}{|c|}{\textcolor{red!99!blue}{990}}&\textcolor{red!1!blue}{10}&\textcolor{black!20!white}{0}&\textcolor{red!100!blue}{1000}\\\cline{3-3}
4&\textcolor{black!20!white}{0}&\textcolor{black!20!white}{0}&\textcolor{black!20!white}{0}&\textcolor{black!20!white}{0}&\textcolor{black!20!white}{0}\\

         \end{tabular}
         \end{minipage}\\\\\hline
    $n=40$&
    $n=100$\\
         \begin{minipage}{0.45\linewidth}
         \centering
       \scriptsize
         \begin{tabular}{c|cccccc|c}
         \backslashbox{g}{m}&3&4&5&6&Total\\\hline
2&\textcolor{red!1.8!blue}{18}&\textcolor{red!1!blue}{10}&\textcolor{black!20!white}{0}&\textcolor{black!20!white}{0}&\textcolor{red!2.8!blue}{28}\\\cline{3-3}
3&\textcolor{black!20!white}{0}&\multicolumn{1}{|c|}{\textcolor{red!95.3!blue}{953}}&\textcolor{red!1!blue}{10}&\textcolor{black!20!white}{0}&\textcolor{red!96.3!blue}{963}\\\cline{3-3}
4&\textcolor{black!20!white}{0}&\textcolor{red!0.9!blue}{9}&\textcolor{black!20!white}{0}&\textcolor{black!20!white}{0}&\textcolor{red!0.9!blue}{9}\\

         \end{tabular}
         \end{minipage}&
         \begin{minipage}{0.45\linewidth}
         \centering
       \scriptsize
         \begin{tabular}{c|cccccc|c}
         \backslashbox{g}{m}&3&4&5&6&Total\\\hline
2&\textcolor{black!20!white}{0}&\textcolor{black!20!white}{0}&\textcolor{black!20!white}{0}&\textcolor{black!20!white}{0}&\textcolor{black!20!white}{0}\\\cline{3-3}
3&\textcolor{black!20!white}{0}&\multicolumn{1}{|c|}{\textcolor{red!99.8!blue}{998}}&\textcolor{red!0.1!blue}{1}&\textcolor{black!20!white}{0}&\textcolor{red!99.9!blue}{999}\\\cline{3-3}
4&\textcolor{black!20!white}{0}&\textcolor{red!0.1!blue}{1}&\textcolor{black!20!white}{0}&\textcolor{black!20!white}{0}&\textcolor{red!0.1!blue}{1}\\

         \end{tabular}
         \end{minipage}\\\\\hline
    $n=60$&
    $n=120$\\
         \begin{minipage}{0.45\linewidth}
         \centering
       \scriptsize
         \begin{tabular}{c|cccccc|c}
         \backslashbox{g}{m}&3&4&5&6&Total\\\hline
2&\textcolor{black!20!white}{0}&\textcolor{red!1!blue}{10}&\textcolor{black!20!white}{0}&\textcolor{black!20!white}{0}&\textcolor{red!1!blue}{10}\\\cline{3-3}
3&\textcolor{black!20!white}{0}&\multicolumn{1}{|c|}{\textcolor{red!97.9!blue}{979}}&\textcolor{red!1.1!blue}{11}&\textcolor{black!20!white}{0}&\textcolor{red!99!blue}{990}\\\cline{3-3}
4&\textcolor{black!20!white}{0}&\textcolor{black!20!white}{0}&\textcolor{black!20!white}{0}&\textcolor{black!20!white}{0}&\textcolor{black!20!white}{0}\\

         \end{tabular}
         \end{minipage}&
         \begin{minipage}{0.45\linewidth}
         \centering
       \scriptsize
         \begin{tabular}{c|cccccc|c}
         \backslashbox{g}{m}&3&4&5&6&Total\\\hline
2&\textcolor{black!20!white}{0}&\textcolor{black!20!white}{0}&\textcolor{black!20!white}{0}&\textcolor{black!20!white}{0}&\textcolor{black!20!white}{0}\\\cline{3-3}
3&\textcolor{black!20!white}{0}&\multicolumn{1}{|c|}{\textcolor{red!99!blue}{990}}&\textcolor{red!0.1!blue}{1}&\textcolor{red!0.9!blue}{9}&\textcolor{red!100!blue}{1000}\\\cline{3-3}
4&\textcolor{black!20!white}{0}&\textcolor{black!20!white}{0}&\textcolor{black!20!white}{0}&\textcolor{black!20!white}{0}&\textcolor{black!20!white}{0}\\

         \end{tabular}
         \end{minipage}\\\\\hline
    \end{tabular}
\end{table}

\begin{table}[h!]
    \centering
    \caption{Distribution of the $(\widehat{g},\widehat{m})$ pairs selected by the proposed procedure with respect to $n$ for $\varepsilon=0.20$. The boxed data corresponds to the reference pair.}
    \null
   \label{tab:Evol20}
  \begin{tabular}{c|c} 
  \multicolumn{2}{c}{$\varepsilon=0.20$}\\
  \hline\hline
   $n=20$&
    $n=80$\\
        \begin{minipage}{0.45\linewidth}
         \centering
        \scriptsize
         \begin{tabular}{c|cccccc|c}
         \backslashbox{g}{m}&2&3&4&5&Total\\\hline
1&\textcolor{red!2.8!blue}{28}&\textcolor{red!0.1!blue}{1}&\textcolor{black!20!white}{0}&\textcolor{black!20!white}{0}&\textcolor{red!2.9!blue}{29}\\
2&\textcolor{red!5.2!blue}{52}&\textcolor{red!48.5!blue}{485}&\textcolor{red!6.7!blue}{67}&\textcolor{black!20!white}{0}&\textcolor{red!60.4!blue}{604}\\\cline{4-4}
3&\textcolor{red!0.1!blue}{1}&\textcolor{red!10.8!blue}{108}&\multicolumn{1}{|c|}{\textcolor{red!23.9!blue}{239}}&\textcolor{black!20!white}{0}&\textcolor{red!34.8!blue}{348}\\\cline{4-4}
4&\textcolor{black!20!white}{0}&\textcolor{red!1.9!blue}{19}&\textcolor{black!20!white}{0}&\textcolor{black!20!white}{0}&\textcolor{red!1.9!blue}{19}\\

         \end{tabular}
         \end{minipage}&
         \begin{minipage}{0.45\linewidth}
         \centering
       \scriptsize
         \begin{tabular}{c|cccccc|c}
         \backslashbox{g}{m}&2&3&4&5&Total\\\hline
1&\textcolor{black!20!white}{0}&\textcolor{black!20!white}{0}&\textcolor{black!20!white}{0}&\textcolor{black!20!white}{0}&\textcolor{black!20!white}{0}\\
2&\textcolor{black!20!white}{0}&\textcolor{black!20!white}{0}&\textcolor{red!3!blue}{30}&\textcolor{black!20!white}{0}&\textcolor{red!3!blue}{30}\\\cline{4-4}
3&\textcolor{black!20!white}{0}&\textcolor{black!20!white}{0}&\multicolumn{1}{|c|}{\textcolor{red!96!blue}{960}}&\textcolor{red!1!blue}{10}&\textcolor{red!97!blue}{970}\\\cline{4-4}
4&\textcolor{black!20!white}{0}&\textcolor{black!20!white}{0}&\textcolor{black!20!white}{0}&\textcolor{black!20!white}{0}&\textcolor{black!20!white}{0}\\

         \end{tabular}
         \end{minipage}\\\\\hline
    $n=40$&
    $n=100$\\
         \begin{minipage}{0.45\linewidth}
         \centering
       \scriptsize
         \begin{tabular}{c|cccccc|c}
         \backslashbox{g}{m}&2&3&4&5&Total\\\hline
1&\textcolor{black!20!white}{0}&\textcolor{black!20!white}{0}&\textcolor{black!20!white}{0}&\textcolor{black!20!white}{0}&\textcolor{black!20!white}{0}\\
2&\textcolor{red!0.1!blue}{1}&\textcolor{red!11.9!blue}{119}&\textcolor{red!7.9!blue}{79}&\textcolor{black!20!white}{0}&\textcolor{red!19.9!blue}{199}\\\cline{4-4}
3&\textcolor{black!20!white}{0}&\textcolor{red!0.9!blue}{9}&\multicolumn{1}{|c|}{\textcolor{red!78.2!blue}{782}}&\textcolor{red!0.9!blue}{9}&\textcolor{red!80!blue}{800}\\\cline{4-4}
4&\textcolor{black!20!white}{0}&\textcolor{black!20!white}{0}&\textcolor{red!0.1!blue}{1}&\textcolor{black!20!white}{0}&\textcolor{red!0.1!blue}{1}\\

         \end{tabular}
         \end{minipage}&
         \begin{minipage}{0.45\linewidth}
         \centering
       \scriptsize
         \begin{tabular}{c|cccccc|c}
         \backslashbox{g}{m}&2&3&4&5&Total\\\hline
1&\textcolor{black!20!white}{0}&\textcolor{black!20!white}{0}&\textcolor{black!20!white}{0}&\textcolor{black!20!white}{0}&\textcolor{black!20!white}{0}\\
2&\textcolor{black!20!white}{0}&\textcolor{black!20!white}{0}&\textcolor{red!0.1!blue}{1}&\textcolor{black!20!white}{0}&\textcolor{red!0.1!blue}{1}\\\cline{4-4}
3&\textcolor{black!20!white}{0}&\textcolor{black!20!white}{0}&\multicolumn{1}{|c|}{\textcolor{red!98!blue}{980}}&\textcolor{red!1.9!blue}{19}&\textcolor{red!99.9!blue}{999}\\\cline{4-4}
4&\textcolor{black!20!white}{0}&\textcolor{black!20!white}{0}&\textcolor{black!20!white}{0}&\textcolor{black!20!white}{0}&\textcolor{black!20!white}{0}\\

         \end{tabular}
         \end{minipage}\\\\\hline
    $n=60$&
    $n=120$\\
         \begin{minipage}{0.45\linewidth}
         \centering
       \scriptsize
         \begin{tabular}{c|cccccc|c}
         \backslashbox{g}{m}&2&3&4&5&Total\\\hline
1&\textcolor{black!20!white}{0}&\textcolor{black!20!white}{0}&\textcolor{black!20!white}{0}&\textcolor{black!20!white}{0}&\textcolor{black!20!white}{0}\\
2&\textcolor{black!20!white}{0}&\textcolor{red!1.1!blue}{11}&\textcolor{red!4.9!blue}{49}&\textcolor{red!0.9!blue}{9}&\textcolor{red!6.9!blue}{69}\\\cline{4-4}
3&\textcolor{black!20!white}{0}&\textcolor{black!20!white}{0}&\multicolumn{1}{|c|}{\textcolor{red!92.8!blue}{928}}&\textcolor{red!0.3!blue}{3}&\textcolor{red!93.1!blue}{931}\\\cline{4-4}
4&\textcolor{black!20!white}{0}&\textcolor{black!20!white}{0}&\textcolor{black!20!white}{0}&\textcolor{black!20!white}{0}&\textcolor{black!20!white}{0}\\

         \end{tabular}
         \end{minipage}&
         \begin{minipage}{0.45\linewidth}
         \centering
       \scriptsize
         \begin{tabular}{c|cccccc|c}
         \backslashbox{g}{m}&2&3&4&5&Total\\\hline
1&\textcolor{black!20!white}{0}&\textcolor{black!20!white}{0}&\textcolor{black!20!white}{0}&\textcolor{black!20!white}{0}&\textcolor{black!20!white}{0}\\
2&\textcolor{black!20!white}{0}&\textcolor{black!20!white}{0}&\textcolor{black!20!white}{0}&\textcolor{black!20!white}{0}&\textcolor{black!20!white}{0}\\\cline{4-4}
3&\textcolor{black!20!white}{0}&\textcolor{black!20!white}{0}&\multicolumn{1}{|c|}{\textcolor{red!98.9!blue}{989}}&\textcolor{red!1!blue}{10}&\textcolor{red!99.9!blue}{999}\\\cline{4-4}
4&\textcolor{black!20!white}{0}&\textcolor{black!20!white}{0}&\textcolor{red!0.1!blue}{1}&\textcolor{black!20!white}{0}&\textcolor{red!0.1!blue}{1}\\

         \end{tabular}
         \end{minipage}\\\\\hline
    \end{tabular}
\end{table}

\begin{table}[h!]
    \centering
    \caption{Distribution of the $(\widehat{g},\widehat{m})$ pairs selected by the proposed procedure with respect to $n$ for $\varepsilon=0.25$. The boxed data corresponds to the reference pair.}
    \null
   \label{tab:Evol25}
  \begin{tabular}{c|c} 
  \multicolumn{2}{c}{$\varepsilon=0.25$}\\
  \hline\hline
    $n=20$&
    $n=80$\\
         \begin{minipage}{0.45\linewidth}
         \centering
       \scriptsize
         \begin{tabular}{c|cccccc|c}
         \backslashbox{g}{m}&2&3&4&5&Total\\\hline
1&\textcolor{red!22.8!blue}{228}&\textcolor{red!0.9!blue}{9}&\textcolor{black!20!white}{0}&\textcolor{black!20!white}{0}&\textcolor{red!23.7!blue}{237}\\
2&\textcolor{red!18.5!blue}{185}&\textcolor{red!44!blue}{440}&\textcolor{red!1!blue}{10}&\textcolor{black!20!white}{0}&\textcolor{red!63.5!blue}{635}\\\cline{4-4}
3&\textcolor{red!1!blue}{10}&\textcolor{red!10.4!blue}{104}&\multicolumn{1}{|c|}{\textcolor{red!1.2!blue}{12}}&\textcolor{black!20!white}{0}&\textcolor{red!12.6!blue}{126}\\\cline{4-4}
4&\textcolor{black!20!white}{0}&\textcolor{black!20!white}{0}&\textcolor{red!0.2!blue}{2}&\textcolor{black!20!white}{0}&\textcolor{red!0.2!blue}{2}\\

         \end{tabular}
         \end{minipage}&
         \begin{minipage}{0.45\linewidth}
         \centering
       \scriptsize
         \begin{tabular}{c|cccccc|c}
         \backslashbox{g}{m}&2&3&4&5&Total\\\hline
1&\textcolor{black!20!white}{0}&\textcolor{black!20!white}{0}&\textcolor{black!20!white}{0}&\textcolor{black!20!white}{0}&\textcolor{black!20!white}{0}\\
2&\textcolor{black!20!white}{0}&\textcolor{red!14.2!blue}{142}&\textcolor{red!13.7!blue}{137}&\textcolor{red!0.9!blue}{9}&\textcolor{red!28.8!blue}{288}\\\cline{4-4}
3&\textcolor{black!20!white}{0}&\textcolor{black!20!white}{0}&\multicolumn{1}{|c|}{\textcolor{red!67.4!blue}{674}}&\textcolor{red!3.8!blue}{38}&\textcolor{red!71.2!blue}{712}\\\cline{4-4}
4&\textcolor{black!20!white}{0}&\textcolor{black!20!white}{0}&\textcolor{black!20!white}{0}&\textcolor{black!20!white}{0}&\textcolor{black!20!white}{0}\\

         \end{tabular}
         \end{minipage}\\\\\hline
    $n=40$&
    $n=100$\\
         \begin{minipage}{0.45\linewidth}
         \centering
       \scriptsize
         \begin{tabular}{c|cccccc|c}
         \backslashbox{g}{m}&2&3&4&5&Total\\\hline
1&\textcolor{red!1.9!blue}{19}&\textcolor{red!1.1!blue}{11}&\textcolor{black!20!white}{0}&\textcolor{black!20!white}{0}&\textcolor{red!3!blue}{30}\\
2&\textcolor{red!1!blue}{10}&\textcolor{red!57.9!blue}{579}&\textcolor{red!9.7!blue}{97}&\textcolor{black!20!white}{0}&\textcolor{red!68.6!blue}{686}\\\cline{4-4}
3&\textcolor{black!20!white}{0}&\textcolor{red!2.1!blue}{21}&\multicolumn{1}{|c|}{\textcolor{red!24.5!blue}{245}}&\textcolor{red!1.8!blue}{18}&\textcolor{red!28.4!blue}{284}\\\cline{4-4}
4&\textcolor{black!20!white}{0}&\textcolor{black!20!white}{0}&\textcolor{black!20!white}{0}&\textcolor{black!20!white}{0}&\textcolor{black!20!white}{0}\\

         \end{tabular}
         \end{minipage}&
         \begin{minipage}{0.45\linewidth}
         \centering
       \scriptsize
         \begin{tabular}{c|cccccc|c}
         \backslashbox{g}{m}&2&3&4&5&Total\\\hline
1&\textcolor{black!20!white}{0}&\textcolor{black!20!white}{0}&\textcolor{black!20!white}{0}&\textcolor{black!20!white}{0}&\textcolor{black!20!white}{0}\\
2&\textcolor{black!20!white}{0}&\textcolor{red!0.9!blue}{9}&\textcolor{red!8!blue}{80}&\textcolor{black!20!white}{0}&\textcolor{red!8.9!blue}{89}\\\cline{4-4}
3&\textcolor{black!20!white}{0}&\textcolor{black!20!white}{0}&\multicolumn{1}{|c|}{\textcolor{red!89.1!blue}{891}}&\textcolor{red!2!blue}{20}&\textcolor{red!91.1!blue}{911}\\\cline{4-4}
4&\textcolor{black!20!white}{0}&\textcolor{black!20!white}{0}&\textcolor{black!20!white}{0}&\textcolor{black!20!white}{0}&\textcolor{black!20!white}{0}\\

         \end{tabular}
         \end{minipage}\\\\\hline
    $n=60$&
    $n=120$\\
         \begin{minipage}{0.45\linewidth}
         \centering
       \scriptsize
         \begin{tabular}{c|cccccc|c}
         \backslashbox{g}{m}&2&3&4&5&Total\\\hline
1&\textcolor{black!20!white}{0}&\textcolor{black!20!white}{0}&\textcolor{black!20!white}{0}&\textcolor{black!20!white}{0}&\textcolor{black!20!white}{0}\\
2&\textcolor{black!20!white}{0}&\textcolor{red!21.7!blue}{217}&\textcolor{red!10.9!blue}{109}&\textcolor{black!20!white}{0}&\textcolor{red!32.6!blue}{326}\\\cline{4-4}
3&\textcolor{black!20!white}{0}&\textcolor{black!20!white}{0}&\multicolumn{1}{|c|}{\textcolor{red!60.6!blue}{606}}&\textcolor{red!5.7!blue}{57}&\textcolor{red!66.3!blue}{663}\\\cline{4-4}
4&\textcolor{black!20!white}{0}&\textcolor{black!20!white}{0}&\textcolor{red!1.1!blue}{11}&\textcolor{black!20!white}{0}&\textcolor{red!1.1!blue}{11}\\

         \end{tabular}
         \end{minipage}&
         \begin{minipage}{0.45\linewidth}
         \centering
       \scriptsize
         \begin{tabular}{c|cccccc|c}
         \backslashbox{g}{m}&2&3&4&5&Total\\\hline
1&\textcolor{black!20!white}{0}&\textcolor{black!20!white}{0}&\textcolor{black!20!white}{0}&\textcolor{black!20!white}{0}&\textcolor{black!20!white}{0}\\
2&\textcolor{black!20!white}{0}&\textcolor{black!20!white}{0}&\textcolor{red!5.8!blue}{58}&\textcolor{red!0.9!blue}{9}&\textcolor{red!6.7!blue}{67}\\\cline{4-4}
3&\textcolor{black!20!white}{0}&\textcolor{black!20!white}{0}&\multicolumn{1}{|c|}{\textcolor{red!92.4!blue}{924}}&\textcolor{red!0.9!blue}{9}&\textcolor{red!93.3!blue}{933}\\\cline{4-4}
4&\textcolor{black!20!white}{0}&\textcolor{black!20!white}{0}&\textcolor{black!20!white}{0}&\textcolor{black!20!white}{0}&\textcolor{black!20!white}{0}\\

         \end{tabular}
         \end{minipage}\\\\\hline
    \end{tabular}
\end{table}

\begin{figure}[ht]
    \centering
    \includegraphics[width=\linewidth]{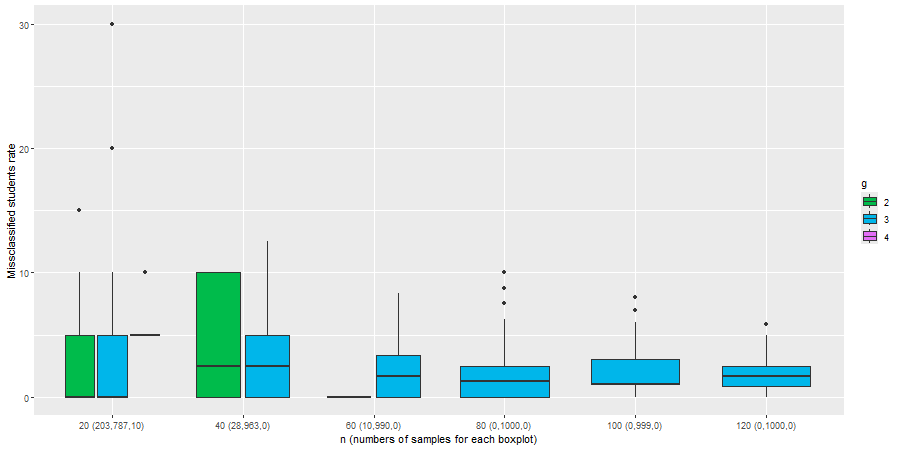}
    \caption{Misclassified students rate distribution with respect to the number $n$ of students and the selected number $\widehat{g}$ of students groups for $\varepsilon=0.15$. Each color corresponds to a given selected number of students groups ($\widehat{g}= 2, 3, 4$). The sample size $n$ is followed by the numbers of samples for each boxplot, when these numbers are greater or equal to 10. For example, the first three boxplots correspond to $n=20$. The left (resp. middle and right) one is drawn from the $203$ (resp. $787$ and $10$) samples from which $\widehat{g}=2$ (resp. $3$ and $4$) groups have been selected.
    }
    \label{fig:boxplot15}
\end{figure}

\begin{figure}[ht]
    \centering
    \includegraphics[width=\linewidth]{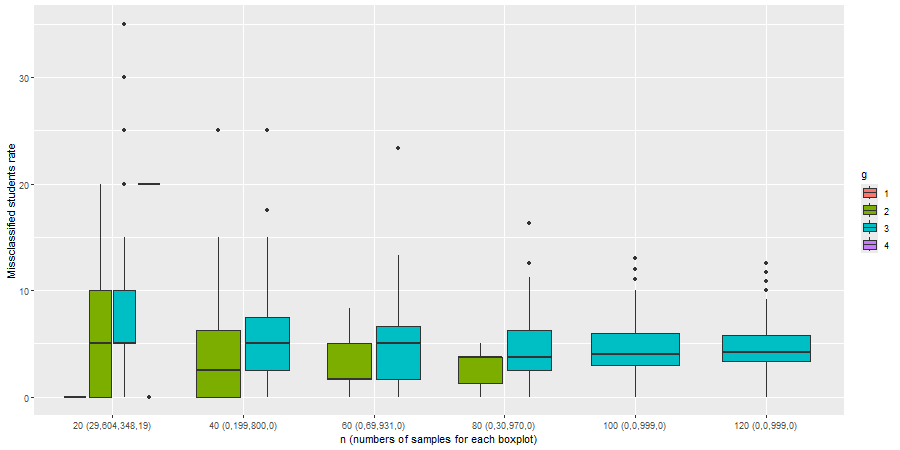}
    \caption{Misclassified students rate distribution with respect to the number $n$ of students and the selected number $\widehat{g}$ of students groups for $\varepsilon=0.2$. Each color corresponds to a given selected number of students groups ($\widehat{g}=1, 2, 3, 4$). The sample size $n$ is followed by the numbers of samples for each boxplot, when these numbers are greater or equal to 10. For example, the first four boxplots correspond to $n=20$. The red (resp. green, blue and purple) one is drawn from the $29$ (resp. $604, 348$ and $19$) samples from which $\widehat{g}=1$ (resp. $2, 3$ and $4$) groups have been selected.
    }
    \label{fig:boxplot20}
\end{figure}

\begin{figure}[ht]
    \centering
    \includegraphics[width=\linewidth]{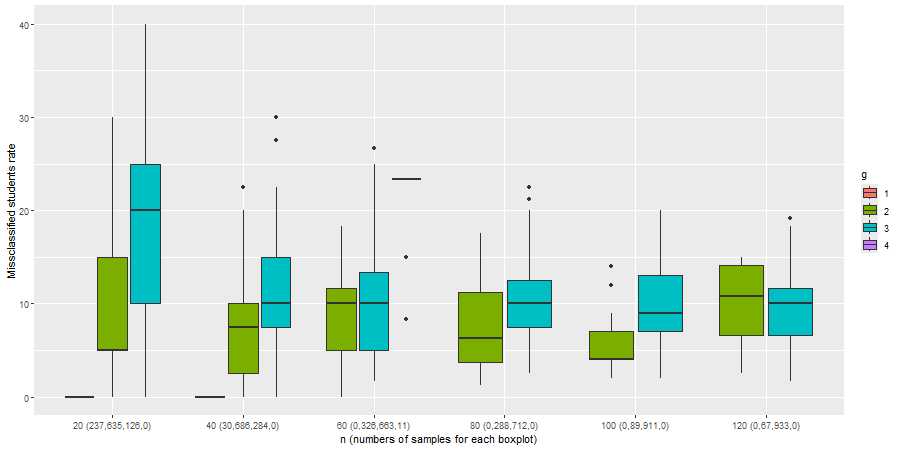}
    \caption{Misclassified students rate distribution with respect to the number $n$ of students and the selected number $\widehat{g}$ of students groups for $\varepsilon=0.25$. Each color corresponds to a given selected number of students groups ($\widehat{g}=1, 2, 3, 4$). The sample size $n$ is followed by the numbers of samples for each boxplot, when these numbers are greater or equal to 10. For example, the first four boxplots correspond to $n=20$. The red (resp. green, blue and purple) one is drawn from the $237$ (resp. $635$, $126$ and $2$) samples from which $\widehat{g}=1$ (resp. $2$, $3$ and $4$) groups have been selected.
    }
    \label{fig:boxplot25}
\end{figure}

\medskip
In Figures ~\ref{fig:boxplot15}, \ref{fig:boxplot20} and \ref{fig:boxplot25}, we focus on the numbers of students groups. The misclassified students rate distribution is displayed with respect to the number of students $n$ and the selected number $\widehat{g}$ of students groups. Note that this rate cannot be greater than $(g-1)/g$ (\cite{robert}). When $\widehat{g}=3$ (the reference number of students groups, blue boxplots), the misclassified students rate decreases when $n$ increases. When $\widehat{g}=2$ (green boxplots), the rate of misclassified students does not tend to decrease but this value of $\widehat{g}$ is less and less selected. Note that values $1$ and $4$ are very rarely selected, whatever the values of $n$ and $\varepsilon$. 

\FloatBarrier
\section{Conclusion}

In this paper, we have proposed a model selection procedure for latent block models on binary data in the context of placement tests. We have examined the effect of initial values in the combined V-Bayes Gibbs algorithm on this procedure. 

\medskip
Our first results show that this effect is depending on the difficulty of the considered case. In simple cases, initial values have very small effects, which implies that only one initialization in the estimation algorithm is sufficient. On the contrary, in more difficult cases such as the English SELF test data set, initial values can have very important effects and lead to an inappropriate model selection. In such cases, we advice to strongly increase the number of estimation algorithm initializations. In the English SELF test data set, although the hyperparameter $a=4$ is chosen in order to avoid empty classes, we have got two small probabilities values $\widehat{\rho_1}=0.137$ and $\widehat{\rho_4}=0.117$. With a smaller value of $a$, the number of initializations might have been reduced but with a higher risk of empty classes (\cite{keribin2015}).     

\medskip
In a second part, we have studied the robustness of our model selection procedure by investigating the selected number of students groups and the stability of the obtained students partitions with respect to the sample size. As expected, the number of selected students groups tends to the reference one as the sample size increases. Moreover, the rate of misclassified students decreases.

\medskip
In the English SELF test data set, we can observe in Table \ref{tab:eng:parameters} one items group for which the probabilities of correct answer are similar whatever the students groups. In such a case, co-clustering methods may fail to provide a meaningful result due to the presence of noisy or irrelevant features. As a perspective, it could be relevant to consider another class of models namely the latent blocks models with noise class proposed by Laclau and Brault \cite{laclau2019noise}. This class of models would enable to include a group of items for which all students would have the same probability of success regardless of the group they belong to. It would make it possible to isolate items which have small impact on the students classification and would speak in favour of the removal of these items in the placement tests.

\section*{Acknowledgment}
This work has been carried out in the framework of the IRS project \texttt{COPOLangues} funded by IDEX Universit\'e Grenoble Alpes. It has also been partially supported by MIAI@Grenoble Alpes (ANR-19-P3IA-0003). All the computations presented in this paper were performed using the GRICAD infrastructure which is supported by Grenoble research communities. The authors thank Sylvain Coulange and Marie-Pierre Jouannaud for the data acquisition and Margaux Leroy for the preliminary robustness study in latent block models with application to an English SELF placement test.

\section*{References}
\bibliographystyle{elsarticle-num} 
\bibliography{Biblio_article_COPOLangues}

\end{document}